\documentclass[aps,prl,floats,floatfix,twocolumn,showpacs,superscriptaddress]{revtex4}

\usepackage{epsfig}
\usepackage{latexsym,amsmath}
\usepackage{amsbsy}

\begin{document}

\title{Navigating ultrasmall worlds in ultrashort time
}

\author{Mari{\'a}n Bogu{\~n}{\'a}}

\affiliation{Departament de F{\'\i}sica Fonamental, Universitat de
  Barcelona, Mart\'{\i} i Franqu\`es 1, 08028 Barcelona, Spain}

\author{Dmitri Krioukov}

\affiliation{Cooperative Association for Internet Data
Analysis (CAIDA), University of California, San Diego (UCSD), 9500
Gilman Drive, La Jolla, CA 92093, USA}

\date{\today}

\begin{abstract}

Random scale-free networks are ultra-small worlds. The average length of
shortest paths in networks of size $N$ scales as $\ln \ln N$. Here
we show that these ultra-small worlds can be navigated in
ultra-short time.
Greedy routing on scale-free networks embedded in metric spaces
finds paths with the average length scaling also as $\ln \ln N$.
Greedy routing uses only local information to navigate a network.
Nevertheless, it finds asymptotically shortest paths,
direct computation of which requires global topology knowledge.
Our findings imply that the peculiar structure of
complex networks insures that the lack of global
topological awareness has asymptotically no impact on the
length of communication paths. These results
have important consequences for communication systems such as the
Internet, where maintaining knowledge of current topology is a major
scalability bottleneck.

\end{abstract}

\pacs{89.75.Hc, 05.45.Df, 64.60.Ak}

\maketitle

Random scale-free networks are ultra-small
worlds~\cite{Cohen:2003,Dorogovtsev:2003b,Chung:2002}. The average
and maximum lengths of shortest paths in scale-free networks with
power-law degree distributions, $P(k) \sim
k^{-\gamma}$, $\gamma\in[2,3]$, scale with network size $N$ as
$\ln\ln N$~\cite{Cohen:2003,Dorogovtsev:2003b}~\footnote{The
diameter in~\cite{Chung:2002} scales as $\ln N$ because as opposed
to~\cite{Cohen:2003,Dorogovtsev:2003b}, to avoid degree correlations
for tractability purposes, the authors impose the maximum-degree
cut-off at $\sim\sqrt{N}$, which is smaller than the natural
maximum-degree cut-off $\sim N^{1/(\gamma-1)}$ in scale-free
networks with $\gamma<3$~\cite{Dorogovtsev:2002}. The average
shortest-path length is still $\sim\ln\ln N$ in~\cite{Chung:2002}.}.
However, finding such shortest paths requires global topology
knowledge, which is not available to nodes in many real networks.
It may seem surprising at first that, having no global
topological awareness, nodes can find any paths to destinations at
all. In~\cite{Boguna:2008} we address this apparent paradox by
showing that the observed topological characteristics of complex networks
maximize their navigability, measured by the efficiency of the greedy
routing process.

Greedy routing (GR)~\cite{Kleinberg:2000,MaNg04,SiJe05,FraGa07,kleinberg06-review}
relies on the hidden
metric space abstraction~\cite{Serrano:2008}.
In this abstraction a network is embedded in a metric space,
with distances in this space representing intrinsic node similarities.
To route information to a given destination, a node forwards
the information to its network neighbor closest to the destination
in this space.
This general mechanism underlies processes ranging from search
in social networks~\cite{TraMi69} to protein folding~\cite{Ravasz:2007}.
The existence of hidden metric spaces under real
networks in general is a conjecture, but we found empirical
evidence of their existence for some real networks,
such as the Internet or some social networks~\cite{Serrano:2008}.
In other cases, the metric space may be visible. In the airport network,
for example, this space is geographic~\cite{ThaAl07,Boguna:2008}.

In~\cite{ThaAl07}, numerical experiments show that
scale-free networks are navigable in a wide region of parameters. Specifically, GR and
its modifications are found to perform generally well, in terms of the length and number
of successful paths, on scale-free
networks embedded in a plane. The GR efficiency is attributed
to network heterogeneity. In~\cite{Boguna:2008}
the analytic results and simulations show that not only
heterogeneity but also clustering affect strongly the GR
efficiency. The thermodynamic limit is considered, and a network is called
navigable if in this limit, GR can find
paths for a macroscopic fraction of source-destination pairs. Navigable networks
are shown to have
sufficiently strong clustering and heterogeneity of node degrees,
i.e., $\gamma\approx2$.

Here we show analytically and in simulations
that the average hop length of paths that GR produces
in these navigable networks scales with network
size as $\bar{\tau} \sim\ln{\ln{N}}/|\ln{(\gamma-2)}|$. Given
that the average length of shortest paths in these networks, as shown
in~\cite{Cohen:2003,Dorogovtsev:2003b}, also scales as
$\mathcal{D}\sim\ln{\ln{N}}/|\ln{(\gamma-2)}|$, we conclude that the
GR paths are asymptotically shortest.

To obtain this result we use the generic class of models introduced
in~\cite{Serrano:2008}. These models generate scale-free networks
embedded in metric spaces as follows. Given a target network size
$N$, first assign to all nodes their coordinates in the metric
space, and an additional hidden variable $\kappa$ representing their
expected degrees. To generate scale-free networks, the variable
$\kappa$ is power-law distributed according to $\rho(\kappa) \propto
\kappa^{-\gamma}$, $\kappa \in [\kappa_{0},\infty)$, where
$\kappa_0$ is the minimum expected degree. The metric space can be
any homogeneous and isotropic $D$-dimensional space. Nodes are
distributed in it with a uniform density $\delta$ that is set to $\delta=1$
without loss of generality. Then each pair of vertices $i$
and $j$ is connected by an edge with probability $r(x)$, $x \equiv
d_{ij}/(\mu \kappa_{i} \kappa_{j})^{1/D}$, where $d_{ij}$ is the
distance between the two vertices in the metric space, and
$\kappa_{i}$ and $\kappa_{j}$ are their expected degrees.

A proper choice of the parameter $\mu$, which depends on a specific
form of the connection probability $r(x)$, guarantees that the
average degree of vertices with hidden variable $\kappa$ is
$\bar{k}(\kappa)=\kappa$, so that $\kappa$ can indeed be identified
with the degree. The exponent $\gamma$ in $\rho(\kappa)$ is then the
power-law exponent of the degree distribution in the resulting
networks~\cite{Serrano:2008}. These properties of the model hold for
any dimension $D$ of the metric space and for any form of the
connection probability $r(x)$, as long as the integral
$\int_{0}^{\infty} x^{D-1}r(x)dx$ is bounded. We thus have a very
versatile class of models since we can independently fix the average
degree and the exponent $\gamma$ without specifying the function
$r(x)$, which can then be used to control clustering in the network.
For example, in~\cite{Serrano:2008} we use $r(x)=(1+x)^{-\alpha}$,
with $\alpha>D$ and
$\mu=\Gamma(D/2)\Gamma(\alpha)/[2\pi^{D/2} \langle k \rangle \Gamma(\alpha-D) \Gamma(D)]$.
This form of $r(x)$ leads to the following two extremes. In the limit
$\alpha \rightarrow D$ clustering vanishes. The network loses its
metric properties and becomes equivalent to a random graph, where the
probability that two nodes are connected depends only on their
expected degrees, and not on the metric distance between them. In
the opposite extreme $\alpha \rightarrow \infty$ clustering
converges to a finite value, and the topology of the network is
strongly influenced by the metric properties of the underlying
space. This latter extreme yields networks lying in the navigable
region.

\begin{figure}[t]
\begin{center}
\epsfig{file=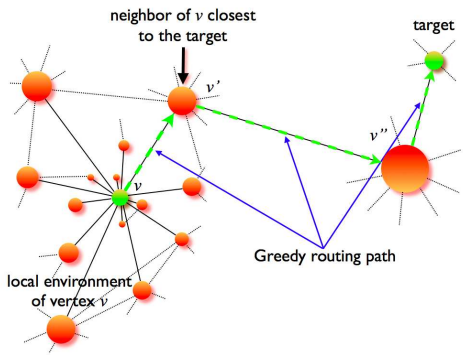,width=6.5cm}
\end{center}
\vspace{-0.5cm} \caption{Illustration of the efficient greedy
routing mechanism. The figure shows the current vertex $v$ with its
local neighborhood. The size of each vertex is proportional to its
degree and the plane represents the underlying metric space. Vertex
$v'$ is the neighbor of $v$ which is closest to the target and also
one of its furthest and highest-degree neighbors. At the next hop, greedy
routing proceeds from $v'$ to $v''$, reaching an even higher-degree
vertex, traveling an even longer distance, and getting much closer to the
target.
}
\label{sketch}
\end{figure}

We next give an intuitive explanation, illustrated in
Fig.~\ref{sketch}, for why GR is efficient in these
navigable networks with strong coupling between network topology and
underlying geometry. Suppose the GR process starts at
some low-degree node and intends to reach a destination located far
away in the metric space. Ideally the process should proceed to
hubs, high-degree nodes, that likely cover long distances by their
numerous connections. However, GR is degree-agnostic, it
checks only underlying distances. Therefore, this ideal scenario
with propagation through the hubs can only be implemented if the
node's neighbor closest to the destination is also its
highest-degree neighbor. But this condition is the more likely
satisfied, the faster $r(x)$ decreases, because the faster $r(x)$ decreases,
the stronger the dependency between a node's degree and the
characteristic scale of distances that the node covers by its connections.
This dependency is simple: the higher the degree of a node, the larger its characteristic
distance scale. (In the non-navigable limit $\alpha \to D$, this
dependency disappears.) Consequently, if the next node along a path has a higher
degree, then the node after the next one has an even higher degree,
and the metric distance between these nodes also increases. On the other
hand, the faster $r(x)$ decreases (e.g., the larger $\alpha$), the stronger
clustering. We thus
see that in the navigable case with strong clustering, GR first travels over a
sequence of nodes with increasing degrees and increasing inter-node
distances. At some point, after the current distance to the
destination becomes comparable to the inter-node distance, this
pattern changes, and the process completes in a finite number of
hops.

We now put this intuition on quantitative grounds. We first compute
the probability that a node of expected degree $\kappa$ has a
neighbor with expected degree $\kappa'$ at a distance $d$ from it,
$P(\kappa',d|\kappa)$. Using results
from~\cite{Boguna:2003b,Serrano:2008}, it is easy to show that this
probability is
\begin{equation}
P(\kappa',d|\kappa)=\frac{\rho(\kappa')}{\mu \langle k \rangle \kappa} d^{D-1}r\left(\frac{d}{(\mu \kappa \kappa')^{\frac{1}{D}}} \right).
\end{equation}
The marginal distribution with respect to $\kappa$ is
\begin{equation}
P(\kappa'|\kappa)=\frac{\kappa'\rho(\kappa')}{\langle k \rangle}
\label{marginal}
\end{equation}
for any function $r(x)$ and any dimension $D$. We next compute the
correlation between variables $\kappa'$ and $d$ conditioned on
$\kappa$. Using Bayes' rule and Eq.~(\ref{marginal}), we write
\begin{equation}
P(d|\kappa,\kappa')=\frac{P(\kappa',d|\kappa)}{P(\kappa'|\kappa)}=\frac{d^{D-1}}{\mu \kappa \kappa'} r\left(\frac{d}{(\mu \kappa \kappa')^{\frac{1}{D}}} \right).
\end{equation}
The average metric distance between two connected vertices with
expected degrees $\kappa$ and $\kappa'$ is then
\begin{equation}
\bar{d}(\kappa,\kappa')=(\mu \kappa \kappa')^{\frac{1}{D}} \int_{0}^{x_{c}} x^D r(x)dx,
\label{average_d}
\end{equation}
where $x_{c}=d_{c}(N) (\mu \kappa \kappa')^{-1/D}$ and $d_{c}(N)$ is
the maximum distance between nodes in the metric space,
$d_{c}(N)\sim N^{1/D}$.

If $r(x)=(1+x)^{-\alpha}$ with $\alpha>D+1$, then the integral in
Eq.~(\ref{average_d}) is bounded and we observe positive
correlations between degrees and distances: the higher the node
degree, the longer the characteristic distances that it covers by
its connections, which is exactly the property guaranteeing
GR efficiency. If $D<\alpha<1+D$, the integral in
Eq.~(\ref{average_d}) diverges and we obtain
\begin{equation}
\bar{d}(\kappa,\kappa') \sim (\mu \kappa \kappa')^{\frac{\alpha}{D}-1} [d_{c}(N)]^{D+1-\alpha}.
\end{equation}
In the limit $\alpha \to D$, $\bar{d}(\kappa,\kappa')$ loses any
dependence on $\kappa$ and $\kappa'$, and becomes a large value
diverging in the thermodynamic limit. As a consequence, degrees and
distances are no longer correlated.
The furthest neighbor no longer tends to have the highest degree. These
arguments explain why the network cannot be navigated
if it loses its metric properties and clustering vanishes.

We now shift our attention entirely to navigable networks with
$\alpha>D+1$ and compute the lengths of greedy paths in them. As the
first step, we calculate the maximum expected degree
$\kappa_{c,nn}(\kappa)$ among all neighbors of a node of
expected degree $\kappa$. In a finite-size network, the variable
$\kappa$ is bounded by a natural cut-off $\kappa_{c}\sim
N^{1/(\gamma-1)}$~\cite{Dorogovtsev:2002}. This cut-off is
calculated as the value of $\kappa=\kappa_{c}$ such that we expect
to find only one node with $\kappa>\kappa_{c}$ out of a sample of
$N$ vertices, $N\int_{\kappa_{c}}\rho(\kappa)d\kappa \sim 1$.
Following the same reasoning, the value of
$\kappa_{c,nn}(\kappa)$ can be evaluated as
\begin{equation}
\kappa\int_{\kappa_{c,nn}(\kappa)}^{\kappa_{c}} P(\kappa'|\kappa) d\kappa' \sim 1,
\end{equation}
which leads to
\begin{equation}
\kappa_{c,nn}(\kappa) \sim \left\{
\begin{array}{lr}
\kappa^{\frac{1}{\gamma-2}} & \kappa<\kappa_{c}^{\gamma-2}\\[0.3cm]
\kappa_{c} & \kappa>\kappa_{c}^{\gamma-2}
\end{array}.
\right.
\label{kcnn}
\end{equation}
This result, together with Eq.~(\ref{average_d}), yields the
following expression for the average distance to the next node along a GR path
from a node of degree $\kappa$
\begin{equation}
\bar{d}_{nn}(\kappa) \sim \kappa^{\frac{\gamma-1}{D(\gamma-2)}} \mbox{\hspace{0.5cm} for \hspace{0.5cm}} \kappa<\kappa_{c}^{\gamma-2}.
\label{dnn}
\end{equation}

\begin{figure}[t]
\begin{center}
\epsfig{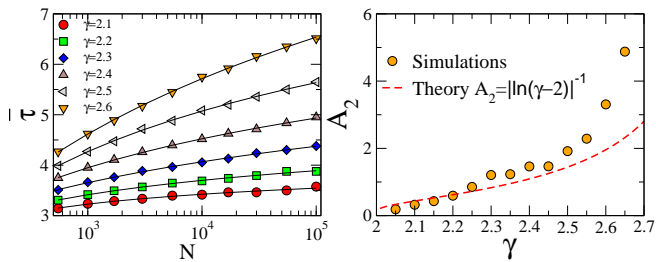}
\end{center}
\vspace{-0.5cm} \caption{Left: average length of GR paths
in generated networks with different values of $\gamma$ as a
function of the system size $N$. Solid lines are fits of the form
$A_{1}+A_{2}\ln{[\ln{N}+A_{3}]}$. Right: parameter $A_{2}$ obtained
from the fit of $\bar{\tau}(N)$ compared to the theoretical
prediction $A_{2}=|\ln{(\gamma-2)}|^{-1}$.} \label{fig1}
\end{figure}

Eqs.~(\ref{kcnn},\ref{dnn}) turn out to be central to our analysis.
First we see from Eq.~(\ref{kcnn}) that only if $1/(\gamma-2)>1$,
i.e., if $\gamma<3$, the degree of the next node along a GR path is, on average,
higher than the degree of the current node. This property explains
why only scale-free networks with $\gamma<3$ are navigable.

Eq.~(\ref{dnn}) also shows that the expected distance between the
next node and the current node of degree $\kappa \sim
N^{(\gamma-2)/(\gamma-1)}$ is $\bar{d}_{nn} \sim
N^{1/D}$, which is of the order of the maximum distance between all
nodes in the metric space. In other words, we can cross the entire
network in a single hop, landing at a node located at a finite and
size-independent distance from the target.
Putting these observations together, we conclude that the time to
reach a target from a low-degree source located far away ($\sim
N^{1/D}$) from the target is roughly the number of hops that it
takes to reach a node of expected degree $\kappa \sim
N^{(\gamma-2)/(\gamma-1)}$, which is a size-dependent contribution,
plus the number of hops needed to cover a finite distance from this
node to the target, which is a
size-independent contribution.

Following these observations, we iterate Eq~(\ref{kcnn})
\begin{equation}
\kappa_{\tau+1} \propto \kappa_{\tau}^{\frac{1}{\gamma-2}} \mbox{\hspace{0.3cm} , \hspace{0.3cm}} \tau=0,1,\cdots
\end{equation}
to find the value of $\tau$ such that $\kappa_{\tau} \sim
N^{(\gamma-2)/(\gamma-1)}$. The solution is
\begin{equation}\label{eq:tau(N)}
\bar{\tau}=A+\frac{\ln{[\ln{N}+B]}}{|\ln{(\gamma-2)}|},
\end{equation}
where $A$ and $B$ are functions of $\gamma$ and $\langle k \rangle$.

This result is remarkable in many respects. First, in the large-size
limit we obtain $\bar{\tau} \sim \ln{\ln{N}}$, meaning that greedy
paths are ultra-short. Second, the prefactor in front of the
logarithm is just a function of $\gamma$, surprisingly independent
of the average degree.
Finally, this prefactor is equal to the prefactor of the average
shortest-path lengths in scale-free
networks~\cite{Cohen:2003,Dorogovtsev:2003b}. It was also shown
in~\cite{Cohen:2003,Dorogovtsev:2003b} that fluctuations around the
average shortest-path lengths are constant. Therefore, in the
thermodynamic limit the shortest-path length distribution becomes a
delta function. This fact, together with the equality between the
average shortest- and greedy-path lengths, implies that for
$N\gg1$ the distribution of greedy-path lengths also converges to
the same delta function. Consequently, in large networks, {\it all greedy
paths are shortest paths}.

To check the accuracy of our theory, we perform extensive numerical
simulations for the model with $D=1$ (a circle) and $\alpha=\infty$,
which is equivalent of taking $r(x)=e^{-x}$ and $\mu=1/(2\langle k
\rangle)$. We also fix the minimum expected degree to
$\kappa_{0}=2$. We note that parameters $\kappa_0$ and $\delta$ are dummy
and can be set to arbitrary values; the only independent
parameters in the model are the average degree $\langle k \rangle$,
the degree exponent $\gamma$, and clustering strength $\alpha$.
Fixing $\kappa_{0}=2$ helps to generate networks
that are fully connected almost surely. If $\kappa_{0}$ is fixed to a
constant, then the average degree depends on $\gamma$ as
$\langle k \rangle=(\gamma-1) \kappa_{0}/(\gamma-2)$. Varying
$\langle k \rangle$ is desirable as it allows us to directly check
with simulations if there is indeed no dependency on $\langle k
\rangle$ of the prefactor in Eq.~(\ref{eq:tau(N)}). We also
verified that networks with a fixed average degree yield the same
results.

Once a network with these parameters is generated, we simulate the
GR process by choosing at random a source and
destination, and forwarding at each node to the node's neighbor
closest to the destination on the circle. The number of
source-destination pairs is $10^6$, and the results are averaged
over a number of network realizations ranging between $400$ and
$4000$. This process is performed for different power-law exponents
$\gamma$ and network sizes $N$. The average GR path
length $\bar{\tau}$ is then computed as a function of $N$ for
different $\gamma$.

The top plot in Fig.~\ref{fig1} shows the results of these
simulations. We then fit empirical $\bar{\tau}(N)$ to a function
of the form $A_{1}+A_{2}\ln{[\ln{N}+A_{3}]}$, where the constants
$A_{1}$, $A_{2}$, and $A_{3}$ are free parameters estimated using
the least square fit to the data. The bottom plot in Fig.~\ref{fig1}
shows the empirical estimate of the coefficient $A_{2}$ compared
with the theoretical prediction $A_{2}=|\ln{(\gamma-2)}|^{-1}$. The
agreement is very good for the values of $\gamma$ close to $2$ and
deteriorates as $\gamma$ approaches $3$. This deterioration is a
consequence of the mean field approximation that assumes that at
each hop the degree increases, which is true only on average. In
fact, for $\gamma$ approaching $3$, there is a increasingly
non-negligible probability of making a hop toward a smaller-degree
node~\cite{Boguna:2008}.
We have also checked~\cite{BoKr09-arxiv} that the fluctuations of
GR path lengths around their average, $\sqrt{\bar{\tau^2}-\bar{\tau}^2}$,
stay constant with increasing $N$, or even slightly
decrease for small $\gamma$.
These observations confirm that in the
thermodynamic limit the GR path length distribution
converges to a delta function.

In summary, we have shown that greedy routing finds asymptotically
shortest paths in scale-free networks with strong clustering and
power-law node degree distribution exponents $\gamma<3$. Given that
topologies of many real networks do have these
properties~\cite{Albert:2002,Newman:2003,Dorogovtsev:2003}, our
findings imply, surprisingly, that even without any global knowledge
of network topology, nodes in complex networks can propagate
information along the shortest routes. In other words, topologies of
many real networks have a peculiar structure that guarantees that
the lack of global topological awareness imposes asymptotically no
impact on the structure of information flows in the network: with or
without the global topology knowledge, information can flow along
the shortest routes. There are other, regular networks, such as lattices,
that also possess these properties, but they require specific embeddings into
specific spaces. Greedy routes on scale-free networks,
on the other hand, are shortest regardless the specifics of a hidden
metric space or connection probability.

Complex networks thus have the structure that allows them
to perform, in the most efficient way, one of their most
basic and common functions: to propagate or signal information to
specific targets through a complex network maze whose global
connectivity is unknown to any node. It remains an open question
if real networks evolve to become navigable~\cite{ClMo03,ChaFra08},
or which networks do have hidden metric spaces underneath and which do not.
Even if such spaces exist, it may be quite challenging to
identify their exact structure. At the same time,
our findings have optimistic practical implications as they open up a possibility to
find shortest-path routing strategies for the Internet that would not require
any global topology knowledge. The requirement for routers to have
and constantly update this knowledge is a major scalability
bottleneck in the Internet today~\cite{iab-raws-report-phys}.

\begin{acknowledgments}
We thank kc claffy and M. \'Angeles Serrano for useful suggestions. This work was supported by
FIS2007-66485-C02-02, Generalitat de Catalunya grant No. SGR00889,
the Ram\'{o}n y Cajal program of the Spanish Ministry of Science,
and by NSF CNS-0434996 and CNS-0722070, DHS N66001-08-C-2029, and
Cisco Systems.
\end{acknowledgments}


\vspace{-0.8cm}
\begin{thebibliography}{22}
\expandafter\ifx\csname natexlab\endcsname\relax\def\natexlab#1{#1}\fi
\expandafter\ifx\csname bibnamefont\endcsname\relax
  \def\bibnamefont#1{#1}\fi
\expandafter\ifx\csname bibfnamefont\endcsname\relax
  \def\bibfnamefont#1{#1}\fi
\expandafter\ifx\csname citenamefont\endcsname\relax
  \def\citenamefont#1{#1}\fi
\expandafter\ifx\csname url\endcsname\relax
  \def\url#1{\texttt{#1}}\fi
\expandafter\ifx\csname urlprefix\endcsname\relax\def\urlprefix{URL }\fi
\providecommand{\bibinfo}[2]{#2}
\providecommand{\eprint}[2][]{\url{#2}}

\bibitem[{\citenamefont{Cohen and Havlin}(2003)}]{Cohen:2003}
\bibinfo{author}{\bibfnamefont{R.}~\bibnamefont{Cohen}} \bibnamefont{and}
  \bibinfo{author}{\bibfnamefont{S.}~\bibnamefont{Havlin}},
  \bibinfo{journal}{Phys. Rev. Lett.} \textbf{\bibinfo{volume}{90}},
  \bibinfo{pages}{058701} (\bibinfo{year}{2003}).

\bibitem[{\citenamefont{Dorogovtsev et~al.}(2003)\citenamefont{Dorogovtsev,
  Mendes, and Samukhin}}]{Dorogovtsev:2003b}
\bibinfo{author}{\bibfnamefont{S.~N.} \bibnamefont{Dorogovtsev}},
  \bibinfo{author}{\bibfnamefont{J.~F.~F.} \bibnamefont{Mendes}},
  \bibnamefont{and} \bibinfo{author}{\bibfnamefont{A.~N.}
  \bibnamefont{Samukhin}}, \bibinfo{journal}{Nucl. Phys. B}
  \textbf{\bibinfo{volume}{653}}, \bibinfo{pages}{307} (\bibinfo{year}{2003}).

\bibitem[{\citenamefont{Chung and Lu}(2002)}]{Chung:2002}
\bibinfo{author}{\bibfnamefont{F.}~\bibnamefont{Chung}} \bibnamefont{and}
  \bibinfo{author}{\bibfnamefont{L.}~\bibnamefont{Lu}}, \bibinfo{journal}{PNAS}
  \textbf{\bibinfo{volume}{99}}, \bibinfo{pages}{15879} (\bibinfo{year}{2002}).

\bibitem[{\citenamefont{Bogu{\~n}\'{a}
  et~al.}(2008)\citenamefont{Bogu{\~n}\'{a}, Krioukov, and
  kc~claffy}}]{Boguna:2008}
\bibinfo{author}{\bibfnamefont{M.}~\bibnamefont{Bogu{\~n}\'{a}}},
  \bibinfo{author}{\bibfnamefont{D.}~\bibnamefont{Krioukov}}, \bibnamefont{and}
  \bibinfo{author}{\bibnamefont{kc~claffy}}, \bibinfo{journal}{Nature Physics,
  doi:10.1038/nphys1130}  (\bibinfo{year}{2008}).

\bibitem[{\citenamefont{Kleinberg}(2000)}]{Kleinberg:2000}
\bibinfo{author}{\bibfnamefont{J.~M.} \bibnamefont{Kleinberg}},
  \bibinfo{journal}{Nature} \textbf{\bibinfo{volume}{406}},
  \bibinfo{pages}{845} (\bibinfo{year}{2000}).

\bibitem[{\citenamefont{Martel and Nguyen}(2004)}]{MaNg04}
\bibinfo{author}{\bibfnamefont{C.}~\bibnamefont{Martel}} \bibnamefont{and}
  \bibinfo{author}{\bibfnamefont{V.}~\bibnamefont{Nguyen}}, in
  \emph{\bibinfo{booktitle}{PODC}} (\bibinfo{year}{2004}).

\bibitem[{\citenamefont{\c{S}im\c{s}ek and Jensen}(2005)}]{SiJe05}
\bibinfo{author}{\bibfnamefont{{\"{O}}.}~\bibnamefont{\c{S}im\c{s}ek}}
  \bibnamefont{and} \bibinfo{author}{\bibfnamefont{D.}~\bibnamefont{Jensen}},
  in \emph{\bibinfo{booktitle}{IJCAI}} (\bibinfo{year}{2005}).

\bibitem[{\citenamefont{Fraigniaud et~al.}(2007)\citenamefont{Fraigniaud,
  Gavoille, Kosowski, Lebhar, and Lotker}}]{FraGa07}
\bibinfo{author}{\bibfnamefont{P.}~\bibnamefont{Fraigniaud}},
  \bibinfo{author}{\bibfnamefont{C.}~\bibnamefont{Gavoille}},
  \bibinfo{author}{\bibfnamefont{A.}~\bibnamefont{Kosowski}},
  \bibinfo{author}{\bibfnamefont{E.}~\bibnamefont{Lebhar}}, \bibnamefont{and}
  \bibinfo{author}{\bibfnamefont{Z.}~\bibnamefont{Lotker}}, in
  \emph{\bibinfo{booktitle}{SPAA}} (\bibinfo{year}{2007}).

\bibitem[{\citenamefont{Kleinberg}(2006)}]{kleinberg06-review}
\bibinfo{author}{\bibfnamefont{J.}~\bibnamefont{Kleinberg}}, in
  \emph{\bibinfo{booktitle}{ICM}} (\bibinfo{year}{2006}).

\bibitem[{\citenamefont{Serrano et~al.}(2008)\citenamefont{Serrano, Krioukov,
  and Bogu{\~n}\'{a}}}]{Serrano:2008}
\bibinfo{author}{\bibfnamefont{M.~A.} \bibnamefont{Serrano}},
  \bibinfo{author}{\bibfnamefont{D.}~\bibnamefont{Krioukov}}, \bibnamefont{and}
  \bibinfo{author}{\bibfnamefont{M.}~\bibnamefont{Bogu{\~n}\'{a}}},
  \bibinfo{journal}{Phys. Rev. Lett.} \textbf{\bibinfo{volume}{100}},
  \bibinfo{pages}{078701} (\bibinfo{year}{2008}).

\bibitem[{\citenamefont{Travers and Milgram}(1969)}]{TraMi69}
\bibinfo{author}{\bibfnamefont{J.}~\bibnamefont{Travers}} \bibnamefont{and}
  \bibinfo{author}{\bibfnamefont{S.}~\bibnamefont{Milgram}},
  \bibinfo{journal}{Sociometry} \textbf{\bibinfo{volume}{32}},
  \bibinfo{pages}{425} (\bibinfo{year}{1969}).

\bibitem[{\citenamefont{Ravasz et~al.}(2007)\citenamefont{Ravasz, Gnanakaran,
  and Toroczkai}}]{Ravasz:2007}
\bibinfo{author}{\bibfnamefont{E.}~\bibnamefont{Ravasz}},
  \bibinfo{author}{\bibfnamefont{S.}~\bibnamefont{Gnanakaran}},
  \bibnamefont{and} \bibinfo{author}{\bibfnamefont{Z.}~\bibnamefont{Toroczkai}}
  (\bibinfo{year}{2007}), \bibinfo{note}{\url{arXiv:0705.0912}}.

\bibitem[{\citenamefont{Thadakamalla et~al.}(2007)\citenamefont{Thadakamalla,
  Albert, and Kumara}}]{ThaAl07}
\bibinfo{author}{\bibfnamefont{H.}~\bibnamefont{Thadakamalla}},
  \bibinfo{author}{\bibfnamefont{R.}~\bibnamefont{Albert}}, \bibnamefont{and}
  \bibinfo{author}{\bibfnamefont{S.}~\bibnamefont{Kumara}},
  \bibinfo{journal}{New J Phys} \textbf{\bibinfo{volume}{9}},
  \bibinfo{pages}{190} (\bibinfo{year}{2007}).

\bibitem[{\citenamefont{Bogu{\~n}\'{a} and
  Pastor-Satorras}(2003)}]{Boguna:2003b}
\bibinfo{author}{\bibfnamefont{M.}~\bibnamefont{Bogu{\~n}\'{a}}}
  \bibnamefont{and}
  \bibinfo{author}{\bibfnamefont{R.}~\bibnamefont{Pastor-Satorras}},
  \bibinfo{journal}{Phys. Rev. E} \textbf{\bibinfo{volume}{68}},
  \bibinfo{pages}{036112} (\bibinfo{year}{2003}).

\bibitem[{\citenamefont{Dorogovtsev and Mendes}(2002)}]{Dorogovtsev:2002}
\bibinfo{author}{\bibfnamefont{S.~N.} \bibnamefont{Dorogovtsev}}
  \bibnamefont{and} \bibinfo{author}{\bibfnamefont{J.~F.~F.}
  \bibnamefont{Mendes}}, \bibinfo{journal}{Adv. Phys.}
  \textbf{\bibinfo{volume}{51}}, \bibinfo{pages}{1079} (\bibinfo{year}{2002}).

\bibitem[{\citenamefont{Bogu{\~n}{\'a} and Krioukov}(2008)}]{BoKr09-arxiv}
\bibinfo{author}{\bibfnamefont{M.}~\bibnamefont{Bogu{\~n}{\'a}}}
  \bibnamefont{and} \bibinfo{author}{\bibfnamefont{D.}~\bibnamefont{Krioukov}}
  (\bibinfo{year}{2008}), \bibinfo{note}{\url{arXiv:0809.2995v1}}.

\bibitem[{\citenamefont{Albert and Barab{\'a}si}(2002)}]{Albert:2002}
\bibinfo{author}{\bibfnamefont{R.}~\bibnamefont{Albert}} \bibnamefont{and}
  \bibinfo{author}{\bibfnamefont{A.-L.} \bibnamefont{Barab{\'a}si}},
  \bibinfo{journal}{Rev. Mod. Phys.} \textbf{\bibinfo{volume}{74}},
  \bibinfo{pages}{47} (\bibinfo{year}{2002}).

\bibitem[{\citenamefont{Newman}(2003)}]{Newman:2003}
\bibinfo{author}{\bibfnamefont{M.~E.~J.} \bibnamefont{Newman}},
  \bibinfo{journal}{SIAM Review} \textbf{\bibinfo{volume}{45}},
  \bibinfo{pages}{167} (\bibinfo{year}{2003}).

\bibitem[{\citenamefont{Dorogovtsev and Mendes}(2003)}]{Dorogovtsev:2003}
\bibinfo{author}{\bibfnamefont{S.~N.} \bibnamefont{Dorogovtsev}}
  \bibnamefont{and} \bibinfo{author}{\bibfnamefont{J.~F.~F.}
  \bibnamefont{Mendes}}, \emph{\bibinfo{title}{Evolution of networks: From
  biological nets to the Internet and WWW}} (\bibinfo{publisher}{Oxford
  University Press}, \bibinfo{address}{Oxford}, \bibinfo{year}{2003}).

\bibitem[{\citenamefont{Clauset and Moore}(2003)}]{ClMo03}
\bibinfo{author}{\bibfnamefont{A.}~\bibnamefont{Clauset}} \bibnamefont{and}
  \bibinfo{author}{\bibfnamefont{C.}~\bibnamefont{Moore}}
  (\bibinfo{year}{2003}), \bibinfo{note}{\url{arXiv:cond-mat/0309415}}.

\bibitem[{\citenamefont{Chaintreau et~al.}(2008)\citenamefont{Chaintreau,
  Fraigniaud, and Lebhar}}]{ChaFra08}
\bibinfo{author}{\bibfnamefont{A.}~\bibnamefont{Chaintreau}},
  \bibinfo{author}{\bibfnamefont{P.}~\bibnamefont{Fraigniaud}},
  \bibnamefont{and} \bibinfo{author}{\bibfnamefont{E.}~\bibnamefont{Lebhar}},
  in \emph{\bibinfo{booktitle}{ICALP}} (\bibinfo{year}{2008}).

\bibitem[{\citenamefont{Meyer et~al.}(2007)\citenamefont{Meyer, Zhang, and
  Fall}}]{iab-raws-report-phys}
\bibinfo{editor}{\bibfnamefont{D.}~\bibnamefont{Meyer}},
  \bibinfo{editor}{\bibfnamefont{L.}~\bibnamefont{Zhang}}, \bibnamefont{and}
  \bibinfo{editor}{\bibfnamefont{K.}~\bibnamefont{Fall}}, eds.,
  \emph{\bibinfo{title}{RFC4984}} (\bibinfo{publisher}{The Internet
  Architecture Board}, \bibinfo{year}{2007}).

\end{thebibliography}
\end{document}